\documentstyle[12pt]{article}

\begin{document}

\pagestyle{myheadings}

\thispagestyle{empty}

{\large\bf\noindent
Relativistic Dynamics of Vector Bosons in the Field of\\
Gravitational Radiation}

{\vskip6mm\hskip16mm\bf
A. Balakin\protect\footnotemark[1]
and V. Kurbanova\protect\footnotemark[1]}

\footnotetext[1]{\footnotesize
\ Department of General Relativity and Gravitation,Kazan State University,
Kremlevskaya 18, Kazan, 420008, Russia;e-mail: dulkyn@mail.ru}

{\vskip5mm\hskip16mm\it Received April 9, 2001}

{\vskip3mm\noindent\small\it
We consider a model of the state evolution of relativistic
vector bosons, which includes both the dynamical equations
for the particle four-velocity  and the equations
for the polarization four-vector evolution in
the field of a nonlinear plane gravitational wave.
In addition to the gravitational minimal coupling,
tidal forces linear in curvature tensor are suggested
to drive the particle state evolution. The exact solutions
of the evolutionary equations are obtained. Birefringence
and tidal deviations from the geodesic motion are discussed.}

\section{\normalsize\bf INTRODUCTION}

The formulation of the covariant dynamic equations for the particles with
internal structure or with supplementary degrees of freedom
unavoidably involves considerati\-on the {\it tidal} forces, i.~e.,
forces linear in the curvature tensor.

A.~Papapetrou in the paper$^{(1)}$, deriving the covariant equations for the
spinning particle, presented the first example of the tidal forces.
Considering the multipole representation of the interaction between the
particle possessing internal structure and the external field,
W.~G.~Dixon$^{(2)}$ and A.~H.~Taub$^{(3)}$ introduced the coupling of the curvature and
the particle quadrupole moment. W.~Israel$^{(4)}$ formulated the covariant dynamics
of the macroscopic polarization in the medium also using the tidal
interactions. At present the covariant theory of the dynamics of spinning
objects is well developed due to the interest to the problem of
emission of the gravitational waves (e.g., Ref.~5).

As it was shown by I.~T.~Drummond and S.~J.~Hathrell in Ref.~6,
the tidal terms appear in the covariant electrodynamics.
The birefringence induced by curvature$^{(6-8)}$ is an electrodynamical phenomenon,
which also admits interpretation in terms of the photon motion under
the influence of tidal forces. This interpretation is based on the fact that
due to gravitationally induced birefringence the energy and the instant
direction of the photon momentum vector depend on the direction of the
polarization vector, rotating in the gravity field.

On the other hand, when the electromagnetic fields dominate and the tidal
forces are negligible in comparison with them, we can use the
relativistic Bargmann-Michel-Telegdi equations$^{(9)}$, describing the minimal
interaction with gravitation and do not displaying the tidal forces.
Nevertheless, as it have been emphasized by I.~B.~Khriplovich$^{(10)}$, there exists a
direct analogy between electromagnetic and tidal coupling of the particle
spin. One can introduce, as was advocated by Khriplovich$^{(10)}$,
new part of the Hamiltonian, namely,  $- \frac{1}{4} R_{abcd} S^{ab} S^{cd}$,
where $R_{abcd}$ is a Riemann curvature tensor and $S^{ab}$ is a spin tensor,
instead of (or in addition to) the electromagnetic coupling part
$\frac{e}{c} F_{ab} S^{ab}$. In other words, both
electromagnetic and tidal coupling of spin can be described analogously. Finally,
I.~B.~Khriplovich noted in Ref.~10 that the developed
formalism and the corresponding equations are applicable to the vector bosons
with the spin equal to one.

We consider this idea as an initial point for our investigations
and we use the evolutionary equations for vector boson with vanishing
electromagnetic field and domina\-ting tidal forces (see the Section~2 of this
paper). In the Section~3 we integrate the dynamic equations exactly for the
case of nonlinear gravitational wave background and present two
particular simplified exact toy-models.

\section{\normalsize\bf EVOLUTIONARY EQUATIONS}

The covariant model of the vector boson evolution contains the pair of
equations.
\begin{equation}
\frac{D{U^{i}}}{D\tau} = - \frac{1}{mc} R^{*i}_{ \ \cdot klm}U^k \Xi^l U^m ,
\label{(1)}
\end{equation}
\begin{equation}
\frac{D{\Xi^{i}}}{D\tau} = - \frac{1}{mc} R^{*i}_{ \ \cdot klm} \Xi^k \Xi^l U^m.
\label{(2)}
\end{equation}
The first equation is a dynamical one for the determination of the particle
four-velocity time-like vector $U^i \equiv \frac{dx^i}{d\tau}$ ($U^i U_i=1$).
The second equation is the equation of evolution of the polarization
space-like four-vector $\Xi^i$ ($\Xi^i\Xi_i = - E^2$, the constant $E$ has
the dimensionality of the Planck constant). $D$ denotes the covariant
differential, $\tau$ is the affine parameter along the particle world-line
(particle proper time). The tensor  $R^{*i}_{ \ \cdot klm}$ is a right-dual to
Riemann tensor
\begin{equation}
R^{*i}_{ \ \cdot klm} \equiv  \frac{1}{2}  R^{i}_{\cdot kpq}
\epsilon^{pq}_{\cdot \cdot lm} .
\label{(3)}
\end{equation}
The Levi-Civita tensor $\epsilon^{pq}_{\cdot \cdot lm}$ is defined as usual
with
\begin{equation}
\epsilon^{pqlm} = \frac{1}{\sqrt{-g}} \ E^{pqlm}, \quad
\epsilon_{pqlm} = \sqrt{-g} \ E_{pqlm},
\label{(4)}
\end{equation}
where $E^{pqlm}$ is a completely skew-symmetric symbol with
$E^{0123} = - E_{0123}= 1$.

The equation (1) looks like the relativistic dynamic equation
$mc\frac{D{U^{i}}}{D\tau}= F^i$ with the force four-vector $F^i$ expressed
in terms of the Riemann tensor. It is a typical "tidal force". This force is
evidently orthogonal to the four-velocity vector ($F^i U_i =0$), and this fact
combined with the velocity normalization law admits us to speak
about particle mass conservation.

Analogously, one can consider the equation (2) as an equation with "tidal
force" acting on the polarization four-vector. The right part of this equation
is apparently orthogonal to the $\Xi^i$ vector. Consequently, $\Xi^k\Xi_k = -
E^2$ is a constant of motion.

Finally, one can see that due to (1) and (2)
\begin{equation}
\frac{D}{D\tau} (U^i \Xi_i) = U_i \frac{D{\Xi^{i}}}{D\tau} +
\Xi_i \frac{D{U^{i}}}{D\tau} \equiv 0 ,
\label{(5)}
\end{equation}
i.e. the scalar product $(U^i \Xi_i) = const$ is an integral of motion.
This constant is considered to be equal to zero in order to display the law of
orthogonality of the four-velocity vector and polarization four-vector for the
massive vector boson as well as for the massless photon$^{(10)}$.

One can note that $S^i$ and $\Xi^i$ quantities are considered to be
pseudo-vectors, neverthe\-less, the product of the right-dual Riemann tensor
with spin or polarization is a true tensor.
For the sake of simplicity we shall omit the prefix "pseudo" below.

\subsection{Remark on the derivation of the evolutionary equations}

The simplest way to obtain the evolutionary equations
(1), (2) is the following. Let us consider the Bargmann-Michel-Telegdi
equations$^{(9)}$ with gyromagnetic ratio $g=2$, i.e. in the case of
the absence of the anomalus magnetic moment:
\begin{equation}
\frac{D{U^{i}}}{D\tau} = \frac{e}{mc^2} F^i_{ \cdot k} \ U^{k}, \quad
\frac{D{S^{i}}}{D\tau} = \frac{e}{mc^2} F^i_{ \cdot k} \ S^{k} .
\label{(6)}
\end{equation}
Then, by the analogy, discussed by I.~B.~Khriplovich$^{(10)}$, let us replace
the  term describing the contribution of the electromagnetic field
$\frac{e}{mc^2} F^i_{ \cdot k}$, by the tidal contribution
$- \frac{1}{2mc} R^{i}_{\cdot klm} S^{lm}$. If we express the
spin-tensor $S^{lm}$ in terms of spin vector
\begin{equation}
S^{lm} = \epsilon^{lmpq} U_p S_q, \quad
S_i = \frac{1}{2}\epsilon_{iklm} U^k S^{lm},
\label{(7)}
\end{equation}
use the definition of the right-dual Riemann tensor (3), and then
replace the spin four-vector $S^i$ by the polarization four-vector
$\Xi^i$, we shall obtain the equations (1) and (2).

\subsection{Gravitational wave background}

The model of the plane gravitational wave (GW) background based on the exact
solution of the Einstein equations in vacuum$^{(11)}$ is well-known and
highly fruitful. The metric of the plane GW can be represented in the
following form:
\begin{equation}
ds^2=2dudv-L(u)^2 \{ [e^{2\beta(u)} (dx^2)^2 + e^{-2\beta(u)} (dx^3)^2 ]\cosh
2\gamma(u)+ 2\sinh 2\gamma(u) dx^2 dx^3 \},
\label{(8)}
\end{equation}
where
\begin{equation}
u\equiv (ct-x^1)/\sqrt{2}, \qquad
v\equiv (ct+x^1)/\sqrt{2}
\label{(9)}
\end{equation}
are the retarded and the advanced time, respectively.
The functions $L(u),  \beta(u), \gamma(u)$ are coupled by one equation
only$^{(11)}$
\begin{equation}
L''(u) +  L (u) \left \{[\beta'(u)]^2 \cosh^2 \gamma(u) + [\gamma'(u)]^2 \right \} = 0,
\label{(10)}
\end{equation}
which is the unique remaining nontrivial Einstein's equation for the
case of vacuum. The prime denotes the derivative by the retarded time $u$.
The functions $\beta(u)$ and $\gamma(u)$ are assumed to be arbitrary ones.
We shall say that the gravitational wave is of the first polarization, if
$\beta(u) \not= 0$, $\gamma(u)=0$, and of the second polarization, if
$\beta(u) = 0$, $\gamma(u) \not=0$.

The initial conditions for the functions $L(u),  \beta(u), \gamma(u)$ are the following:
\begin{equation}
L(0)=1, \quad L'(0)=0,
\quad \beta(0)=\gamma(0)=0,
\quad \beta'(0)=\gamma'(0)=0.
\label{(11)}
\end{equation}
The Riemann tensor calculated with the metric (8) has only four nontrivial
components
$$
-R_{\cdot u2u}^2=R_{\cdot u3u}^3=L^{-2}[L^2 \beta'\cosh^2{2\gamma}]',
$$
$$
R_{\cdot u2u}^3=-L^{-2}[L^2e^{2\beta }(\gamma' -
\beta'\sinh2\gamma \cosh2\gamma )]',
$$
\begin{equation}
R_{\cdot u3u}^2=-L^{-2}[L^2e^{-2\beta }(\gamma' +
\beta' \sinh2\gamma \cosh2\gamma )]'.
\label{(12)}
\end{equation}
The components of the right-dual Riemann tensor are equal, correspondingly, to
$$
R^{*}_{2u2u}= - L^2 \cdot R^{3}_{\cdot u2u} , \quad
R^{*}_{3u3u}= - L^2 \cdot R^{2}_{\cdot u3u} ,
$$
\begin{equation}
R^{*}_{2u3u} = R^{*}_{3u2u} =
L^2 \cdot R_{\cdot u2u}^2 = - L^2 \cdot R_{\cdot u3u}^3 \ .
\label{(13)}
\end{equation}
The space-time with metric (8) admits the existence of five Killing's vectors,
three of them, namely,
\begin{equation}
\xi_{(v)}^i = \delta^i_v ,\quad
\xi_{(2)}^i = \delta^i_2 ,\quad
\xi_{(3)}^i = \delta^i_3 ,
\label{(14)}
\end{equation}
form the Abelian subgroup.
The vector $\xi_{(v)}^i$ possesses the following properties:
\begin{equation}
g_{ik} \xi_{(v)}^i \xi_{(v)}^k = 0 , \quad
\nabla_k \xi_{(v)}^i  = 0 , \quad
\xi_{(v)}^i R_{iklm} = 0 = \xi_{(v)}^i R^*_{iklm},
\label{(15)}
\end{equation}
i.e., it is a null covariantly constant vector, orthogonal
to the Riemann tensor and to its dual tensors.

\section{\normalsize\bf THE EXACT SOLUTION FOR THE MODEL DYNAMICAL SYSTEM}

\subsection{The first integrals of motion}

We have found (see Section~2) that due to the structure of the evolutionary
equations three quadratic quantities
\begin{equation}
U^i U_i = const \equiv 1 , \quad
\Xi_i \Xi ^i = const \equiv - E^2, \quad
\Xi_i U^i = const \equiv 0
\label{(16)}
\end{equation}
are constant along the particle world-line independently on the gravitational
background properties.

Next two integrals of motion
\begin{equation}
U_v \equiv = g_{ik} \ \xi^i_{(v)} U^k = const \equiv C_v , \quad
\Xi_v \equiv = g_{ik} \ \xi^i_{(v)} \Xi^k = const \equiv E_v
\label{(17)}
\end{equation}
exist due to the GW symmetry. The convolution of the equations (1) and (2)
with $g_{ik}\xi^k_{(v)}$ gives the formulae (17), if we use the properties (15)
of the null covariantly constant Killing vector.

Finally, accounting that
\begin{equation}
\frac{du}{d\tau} = U^u = U_v = C_v ,
\label{(18)}
\end{equation}
we can link the particle proper time $\tau$ with the retarded time $u$:
\begin{equation}
\tau = \frac{u}{C_v} + const .
\label{(19)}
\end{equation}
Note, that for massive boson $C_v \neq 0$.
We shall use the relationship (19) with the constant equal to zero in order to
reparametrize the remaining differential equations for the components of
$U^i$ and $\Xi^i$, parallel to the  GW front plane.

\subsection{The motion in the GW front plane}

Let us extract four equations from (1) and (2) by means of
convolution of (1),(2) with the second and the third Killing vectors from the
abelian subgroup (14).
Denoting the corresponding convolutions by
\begin{equation}
U_\alpha \equiv g_{ik} \ \xi^i_{(\alpha)} U^k , \quad
\Xi_\alpha \equiv g_{ik} \ \xi^i_{(\alpha)} \Xi^k ,
\label{(20)}
\end{equation}
where the Greek indices run from 2 to 3, we obtain the following system of
coupled equations:
\begin{equation}
U_\alpha'(u) = - \frac{1}{mc} \ R^{*}_{\alpha u \gamma u}
g^{\gamma \beta} [C_v \Xi_\beta(u) - E_v U_\beta(u)],
\label{(21)}
\end{equation}
\begin{equation}
\Xi_\alpha'(u) = \frac{1}{C_v}
\left[ \frac{1}{2} g'_{\alpha \gamma}(u)
- \frac{E_v}{mc} \ R^{*}_{\alpha u \gamma u} \right] g^{\gamma \beta}
[C_v \Xi_\beta(u) - E_v U_\beta(u)].
\label{(22)}
\end{equation}
The evident symmetry of the equations (21),(22)
allows us to introduce some new unknown functions
\begin{equation}
X_\alpha(u) \equiv C_v \Xi_\alpha (u)  - E_v U_\alpha (u).
\label{(23)}
\end{equation}
Then we obtain the following two-dimensional {\it key subsystem} as the
differential consequen\-ce of (21),(22):
\begin{equation}
\left( \begin{array}{c} X_2' \\ X_3' \end{array} \right) =
\left(\begin{array}{cc}
a_2^2 & a_2^3 \\
a_3^2 & a_3^3
\end{array}\right)
\left(\begin{array}{c} X_2 \\ X_3 \end{array} \right).
\label{(24)}
\end{equation}
Here
$$
a_2^2(u) = \frac{L'}{L} + \beta'\cosh^2{2\gamma}, \quad
a_2^3(u) = e^{2\beta} [\gamma' - \beta'\sinh2\gamma \cosh2\gamma ],
$$
\begin{equation}
a_3^2(u) = e^{-2\beta}[\gamma' + \beta' \sinh2\gamma \cosh2\gamma], \quad
a_3^3(u) = \frac{L'}{L} - \beta' \cosh^2{2\gamma}.
\label{(25)}
\end{equation}
It is interesting to mention that the key subsystem happens to be self-closed
and does not contain the tidal terms with the Riemann tensor.
Such two-dimensional subsystems were considered,
solved and used in Refs.~8 and~12.
The solution of (24), (25) is the following
\begin{equation}
\left( \begin{array}{c} X_2 \\ X_3 \end{array} \right) =
L \left(\begin{array}{cc}e^{\beta } & 0 \\ 0 & e^{-\beta}\end{array}\right)
\left(\begin{array}{cc} \cosh \gamma  & \sinh \gamma  \\ \sinh \gamma  & \cosh \gamma \end{array}\right)
\left(\begin{array}{cc}
\cos \psi  & -\sin \psi  \\ \sin \psi  & \cos \psi \end{array} \right)
\left( \begin{array}{c} X_2(0) \\ X_3(0) \end{array} \right),
\label{(26)}
\end{equation}
where the function $\psi(u)$ has the form
\begin{equation}
\psi(u) \equiv  \int\limits_0^u \beta' \sinh2\gamma du .
\label{(27)}
\end{equation}
The initial data $X_\alpha(0)$ are predetermined by those for the four-velocity
and polarization vectors: \begin{equation}
X_\alpha(0) = E_\alpha C_v - E_v C_\alpha, \quad
C_\alpha \equiv U_\alpha (0) , \quad E_\alpha \equiv \Xi_\alpha(0) .
\label{(28)}
\end{equation}
Then the solution of the total four-dimensional subsystem of equations
(21),(22) can be reduced to the quadratures
\begin{equation}
\Xi_{\alpha}(u) = \frac{1}{C_v} \left[ X_{\alpha}(u) + E_v U_{\alpha}(u)
\right],
\label{(29)}
\end{equation}
\begin{equation}
U_\alpha(u) = C_\alpha  - \frac{1}{mc} \int\limits_0^u du
 \ R^{*}_{\alpha u \gamma u}(u) g^{\gamma \beta}(u) X_\beta(u).
\label{(30)}
\end{equation}
In (8),(12),(13) and (26) one can find all the functions which are
necessary for the integration in (30).

\subsection{The quadratures for the remaining unknown functions}

Now, when $X_\alpha$ are obtained and $U_\alpha(u)$ as well as $\Xi_\alpha(u)$
are represented via $X_\alpha$, we can extract $U_u$ and $X_u$ from the
quadratic integrals (16):
\begin{equation}
U_u(u) = \frac{1}{2C_v} [ 1 - g^{\alpha \sigma}(u) U_\alpha(u) U_\sigma(u)],
\qquad C_v \neq 0,
\label{(31)}
\end{equation}
\begin{equation}
\Xi_u (u) = - \frac{1}{2E_v} [  E^2 + g^{\alpha \sigma}(u)
\Xi_\alpha(u) \Xi_\sigma(u)],
\qquad E_v \neq 0.
\label{(32)}
\end{equation}
Finally, the particle position as a function of the retarded time is described
by the formulae
\begin{equation}
x^\alpha (u) = x^\alpha (0)  + \frac{1}{C_v}\int\limits_0^u g^{\alpha \beta}(u)
U_{\beta}(u) du,
\label{(33)}
\end{equation}
\begin{equation}
v (u) = v(0) + \frac{1}{C_v}\int\limits_0^u  U_u(u) du .
\label{(34)}
\end{equation}
Thus, the formulae (17),(19),(29)-(34) represent in the quadratures the {\it
exact solution} of the equations (1),(2), containing 12 functions of the
proper time $\tau$.

\subsection{The first exact toy-model}

If the massive boson was at rest at the initial moment $u=0$ in the chosen
reference frame, i.e.
\begin{equation}
C_2 = C_3 = 0, \quad C_v = \frac{1}{\sqrt{2}} = U_u(0),
\label{(35)}
\end{equation}
and if the GW has only the first polarization, i.e. $\gamma(u)=0$, then
$\psi(u) \equiv 0$ and we can simplify the formulae (26),(30),(29):
\begin{equation}
X_2(u) = \frac{1}{\sqrt{2}} L e^{\beta}  E_2, \quad
X_3(u) = \frac{1}{\sqrt{2}} L e^{-\beta} E_3, \quad
g^{\alpha \beta} X_\alpha X_\beta = - \frac{1}{2} (E^2_2 + E^2_3),
\label{(36)}
\end{equation}
\begin{equation}
U_2(u) =  - \frac{E_3}{\sqrt{2}mc} \left(L e^{\beta}\right)', \quad
U_3(u) =    \frac{E_2}{\sqrt{2}mc} \left(L e^{-\beta}\right)',
\label{(37)}
\end{equation}
\begin{equation}
\Xi_2(u) =  L e^{\beta} E_2 - \frac{E_v E_3}{mc} \left(L e^{\beta}\right)' , \quad
\Xi_3(u) =  L e^{-\beta} E_3 + \frac{E_v E_2}{mc} \left(L e^{-\beta}\right)' .
\label{(38)}
\end{equation}
The relationships (16) yield
\begin{equation}
2 E^2_v + E^2_2 + E^2_3 = E^2 ,
\label{(39)}
\end{equation}
i.e. one of the constants $E_v$, $E_2$ or $E_3$ can be expressed in terms of
other ones and of the $E$ normalization constant.

Computing the particle energy
\begin{equation}
{\cal E}(u) \equiv mc^2 U_0 = \frac{mc^2}{\sqrt{2}} (U_u + U_v) =
\frac{mc^2}{2\sqrt{2}C_v} ( 1 - g^{\alpha\beta} U_\alpha U_\beta + 2C^2_v),
\label{(40)}
\end{equation}
and using (37) we obtain
\begin{equation}
{\cal E}(u) -  mc^2 = \frac{1}{4 m} \left[ E^2_3 \left(\frac{L'}{L} +
\beta' \right)^2 + E^2_2 \left(\frac{L'}{L} - \beta' \right)^2  \right] .
\label{(41)}
\end{equation}
Finally, let us calculate the $U_1$ component of the four-velocity vector:
\begin{equation}
U_1(u) \equiv \frac{1}{\sqrt{2}}(U_v - U_u) = - \frac{1}{4 m^2 c^2}
\left[ E^2_3 \left(\frac{L'}{L} + \beta' \right)^2 + E^2_2 \left(\frac{L'}{L}
- \beta' \right)^2  \right] .
\label{(42)}
\end{equation}

\subsection{The second toy-model}

Let the GW be of the first polarization as well. Let at
the moment $u=0$ the polarization three-vector of a particle
be directed along the $0x^1$ axis, and the initial
particle motion be one-dimensional, for example along $0x^2$ axis.
It is possible, if the constants are chosen as follows:
\begin{equation}
E_2 = E_3 = 0 , \quad
E_v = - \Xi_u(0) = \frac{E}{\sqrt{2}} , \quad
C_3 = 0 , \quad
C_v = U_u(0) = \sqrt{\frac{1 + C^2_2}{2}}.
\label{(43)}
\end{equation}
Then we see that
\begin{equation}
U_2(u) = C_2 , \quad
U_3(u) =  - \frac{C_2 E}{\sqrt{2}mc} \left(L e^{-\beta}\right)' ,
\label{(44)}
\end{equation}
\begin{equation}
U_1(u) =  \frac{C^2_2}{2\sqrt{1 + C^2_2}} \left[
\left( 1 - \frac{e^{-2\beta}}{L^2}\right) -
\frac{E^2}{2 m^2 c^2 } \left(\frac{L'}{L} - \beta' \right)^2 \right] ,
\label{(45)}
\end{equation}
\begin{equation}
\Xi_2(u) =  \frac{E C_2}{\sqrt{1 + C^2_2}} \left( 1 - L e^{\beta} \right) ,
\quad
\Xi_3(u) =  \frac{E U_3}{\sqrt{1 + C^2_2}}.
\label{(46)}
\end{equation}
The formulae of two  last subsections will simplify the next discussion.

\section{\normalsize\bf DISCUSSION}

The obtained exact solution of the evolutionary model demonstrates
explicitly the following general properties.

(i) {\it Birefringence induced by curvature}

Using the formulae (40) with (30) or an explicit formula (42) we can conclude
that the particle energy at the moment $u$ depends on the initial data for the
polarization four-vector. Since the polarization four-vector is normalized and
orthogonal to the four-velocity (see (16)), only two initial components of this
four-vector are  independent (see (39)). In general case it is convenient to
choose the $E_2$ and $E_3$ parameters to be independent ones. Since the
particle energy depends on two  initial polarization parameters, we deal with
the bosonic analogue of the optical {\it birefringence}.
Since the polarization parameters are involved into the particle energy
formula just due to the tidal interaction, we can denote the effect of this
type as birefringence induced by curvature$^{(6-8)}$.
We can compute the energy shift (41) for $E_2=0$ and for $E_3=0$.
The corresponding ratio of the polarizationally induced energy shifts
is equal to
\begin{equation}
[{\cal E} -  mc^2]_{\vert E_2=0}  :  [{\cal E} - mc^2]_{\vert E_3=0}  =
(L'+L\beta')^2 : (L'-L\beta')^2.
\label{(47)}
\end{equation}
Note that the polarizationally induced energy shifts for the first toy-model
are of the same sign. In the weak GW field they differ only in the second order
of $\beta$. Nevertheless, in the case of
initially moving particle these shifts differ more considerably
and have contributions linear in $\beta$.

The second toy-model demonstrates the degenerate case: if the $E_2$
and $E_3$ compo\-nents vanish, i.e. the initial direction of the polarization
three-vector coincides with that of the GW propagation, then the energy
depends only on the polarization four-vector modulus, and birefringence is
not displayed.

(ii) {\it Particle nongeodesic motion}

In the presence of the tidal interaction the particle motion is no longer the
geodesic one. The tidal force in the right part of the formula (1)
produces the particle {\it acceleration} and the  {\it rotation} of the
four-velocity vector. The formulae (30),(31),(26) describe explicitly
these phenomena. Speaking about the particle rotation,
we emphasize the following feature.
If the initial direction of the particle motion is along
the $0x^2$ axis (see the second toy - model), then in the field of tidal
forces the $U_2$ component remains constant, but the $U_3(u)$ and $U_1(u)$
components appear. Certainly, it is not a standard rotation, because the
instantaneous radius of the quasi-orbit grows. But one can find some analogy
with the rotation of the charged particle in the nonhomogeneous magnetic
field.

In principle, the phenomena of the particle nongeodesic acceleration and
rotation are the supplementary contribution
to the geodesic ones, caused by the minimal coupling with gravitation.
This statement can be illustrated by the
formula (45), in which the term in the parentheses describes the geodesic
variation of the longitudinal velocity, and the last term appears due to the
tidal interactions.

(iii) {\it Rotation of the polarization vector}

The effect of the rotation of the polarization vector is displayed by the
formulae (29) and (26). The third two-dimensional matrix in the product
in the right part of (26) is the standard rotation matrix,
the $\psi(u)$ being the phase of rotation.
Even if we shall neglect the term with right-dual Riemann
curvature tensor in (30), nevertheless, the polarization four-vector (see
(29)) will rotate due to the properties of the $X_\alpha(u)$ (26). In
addition, the rotation of the particle three-velocity produces the
supplementary rotation of the polarization three-vector, which is induced
originally by curvature.

Thus, we can conclude that the gravitational waves effect  on the vector
boson can be considered as a combination of geodesic and tidal acceleration
and rotation, leading to the birefringence phenomenon.

\subsection*{ACKNOWLEDGEMENTS}

The authors are grateful to  W.~Zimdahl and C.~L\"ammerzahl for the fruitful
discussions. This work was supported by the Deutsche Forschungsgemeinschaft.

\subsection*{REFERENCES}

1. A.~Papapetrou, {\it Proc.Roy.Soc.}, A{\bf 209}, p.~248 (1951).\\
2. W.~G.~Dixon, {\it Phil.Trans.Roy.Soc.London}, A{\bf 277}, p.~59 (1974).\\
3. A.~H.~Taub, {\it J. Math. Phys.}, {\bf 5}, p.~112 (1964).\\
4. W.~Israel, {\it GRG}, {\bf 9}, No 5, 451-468 (1978).\\
5. C.~L\"ammerzahl, C.~W.~F.~Everitt, and F.~W.~Hehl (eds.):
{\it Gyros, Clocks, and Interferometers:
Testing Relativistic Gravity in Space}
(Springer-Verlag, Berlin, 2000).\\
6. I.~T.~Drummond and S.~J.~Hathrell, {\it Phys. Rev. D.},
{\bf 22}, 343-395 (1980).\\
7. R.~Lafrance and R.~C.~Myers, {\it Phys. Rev. D.},
{\bf 51}, 2584-2590 (1995).\\
8. A.~B.~Balakin, {\it Clas.Quantum Grav.}, {\bf 14}, 2881-2893 (1997).
9. V.~Bargmann, L.~Michel, and V.~L.~Telegdi, {\it Phys.Rev.Let.},
{\bf 2}, p.~435, (1959).\\
10. I.~B.~Khriplovich, {\it Sov.Phys. JETP}, {\bf 69}(2), 217-219, (1989).\\
11. D.~Kramer, H.~Stephani, M.~McCallum, E.~Herlt, and E.~Schmutzer,
{\it Exact solutions of the Einstein field equations.}  (Berlin, 1980).\\
12. V.~R.~Kurbanova and A.~B.~Balakin, in {\it Recent problems in
field theory. 1998.}, A.~V.~Aminova ed., 211-216 (Kantsler, Kazan, 1998).

\end{document}